%% file: dummy_detection.tex
\documentclass[conference]{IEEEtran}
\IEEEoverridecommandlockouts
\usepackage[nosort,nospace]{cite}
\usepackage{amsmath,amssymb,amsfonts}
\usepackage{algorithmic}
\usepackage{graphicx}
\usepackage{textcomp}
\usepackage{xcolor}
\usepackage{amssymb,amsmath,amsfonts,amsthm}
\usepackage{mathrsfs}
\usepackage{cite}
\usepackage{array}
\usepackage{tabularx}
\usepackage{color}
\usepackage{bm}
\usepackage{mathtools}
\usepackage{subfigure}
\usepackage{mathtools}

\input{commands}

\def\BibTeX{{\rm B\kern-.05em{\sc i\kern-.025em b}\kern-.08em
    T\kern-.1667em\lower.7ex\hbox{E}\kern-.125emX}}
\begin{document}

\title{Detecting Active Attacks in Over-the-Air Computation using Dummy Samples}

\author{David Nordlund, Zheng Chen and Erik G. Larsson\\
	Department of Electrical Engineering, Link\"{o}ping University, Sweden.\\ Email: \{david.nordlund, zheng.chen, erik.g.larsson\}@liu.se
	\thanks{This work was supported in part by Zenith, ELLIIT, Swedish Research Council (VR), Swedish Foundation for Strategic Research (SSF)-SURPRISE, and Security Link.}}

\maketitle

\input{abstract}


\input{introduction}

\input{OtA}
\input{system_model}
\input{results}
\input{conclusions}

\bibliographystyle{IEEEtran}
\bibliography{bib}

\end{document}

%% file: commands.tex
\newcommand{\reals}{\mathbb{R}}
\newcommand{\complexes}{\mathbb{C}}

\newcommand{\norm}[1]{\left\lVert #1 \right\rVert}
\newcommand{\abs}[1]{\left|#1\right|}
\newcommand{\CN}[2]{\mathcal{CN}\left( #1,#2 \right)}
\newcommand{\N}[2]{\mathcal{N}\left( #1,#2 \right)}

\newcommand{\gldec}[2]{\underset{#1}{\overset{#2}{\gtrless}}}
\newcommand{\decrule}{\gldec{\mathcal{H}_0}{\mathcal{H}_1}}
\newcommand{\vect}[1]{\textbf{#1}}

\newcommand{\numusers}{\ensuremath{K}}
\newcommand{\noise}{\ensuremath{\mathbf{z}}}
\newcommand{\cnvar}{\sigma^2}

\newcommand{\dnvar}{\sigma_d^2}
\newcommand{\channelcoef}{\ensuremath{h}}
\newcommand{\psf}{\eta}

\newcommand{\datavector}{\ensuremath{\mathbf{x}}}
\newcommand{\ones}{\ensuremath{\mathbf{u}}}

\newcommand{\byzdatavector}{\ensuremath{\mathbf{b}}}
\newcommand{\numreal}{\ensuremath{L}}
\newcommand{\numdummy}{\ensuremath{D}}

\usepackage[colorinlistoftodos,prependcaption,backgroundcolor=black!5!white,bordercolor=red]{todonotes}

\newcommand{\EX}[1]{\mathbb{E}\{#1\}}
\newcommand{\Var}[1]{\textnormal{Var}\{#1\}}

%% file: abstract.tex
\begin{abstract}
Over-the-Air (OtA) computation is a newly emerged concept for computing functions of data from distributed nodes by taking advantage of the wave superposition property of wireless channels. Despite its advantage in communication efficiency, OtA computation is associated with significant security and privacy concerns that have so far not been thoroughly investigated, especially in the case of active attacks. In this paper, we propose and evaluate a detection scheme against active attacks in OtA computation systems. More explicitly, we consider an active attacker which is an external node sending random or misleading data to alter the aggregated data received by the server. To detect the presence of the attacker, in every communication period, legitimate users send some dummy samples in addition to the real data. We propose a detector design that relies on the existence of a shared secret only known by the legitimate users and the server, that can be used to hide the transmitted signal in a secret subspace. After the server projects the received vector back to the original subspace, the dummy samples can be used to detect active attacks. We show that this design achieves good detection performance for a small cost in terms of channel resources. 
\end{abstract}

%% file: introduction.tex
\section{Introduction}

Motivated by the Internet-of-Things paradigm, Over-the-Air (OtA) computation has emerged as a communication-efficient framework to compute functions of data across multiple source nodes over wireless channels \cite{ota-old-new, sahin2023survey}. Specifically, this is achieved by allowing all source nodes to transmit simultaneously sharing the same frequency/time resources such that the transmitted signals are directly superimposed in the air and received at one common server/fusion center. With proper processing of signals at the source nodes and the server, a variety of functions can be computed/estimated. As compared to traditional designs with orthogonal division of resources, OtA computation shows a clear advantage in terms of system scalability and resource efficiency, especially when error-free communication is not required \cite{Nazer2007ComputationOM,zhu2021over,wang2022overtheair}.  

Despite its advantages, OtA computation is associated with significant security and privacy concerns. These concerns may be related to different types of attacks, including passive attacks with nodes eavesdropping the signals and inferring information about the transmitted data, and active attacks with nodes transmitting perturbation signals to disturb the aggregation process.
Several existing works have considered passive attacks with eavesdroppers. In \cite{Frey2021passive}, a \textit{friendly jammer} is introduced in the transmission design so that the jamming signals can be compensated by the server while impeding the eavesdropper's ability to infer information. In \cite{massny2023secure}, zero-forced artificial noise perturbations are added to each user's data vector before the transmission. The idea is that these perturbation signals sum to zero at the server, but not at the eavesdropper due to the mismatch between the channels of the legitimate users and of the eavesdropper. A similar design with correlated additive perturbations has also been adopted in the area of OtA federated learning (FL) \cite{liao2022passive}.

Compared to passive attacks, there is relatively less research on active attacks in OtA computation systems. Here we define active attacks as the existence of one or more external or legitimate user(s) sending random or misleading data to alter the received aggregated result. In the context of distributed learning with OtA gradient aggregation, the effect of active attacks (generally referred to as Byzantine attacks) can be mitigated by using hierarchical architectures with robust gradient aggregation rules \cite{sifaou2022robust}. However, such techniques are not applicable directly in pure OtA computation, where the goal is to compute functions of distributed data, without specifying how the computation result will be used in specific applications.
On another track, jamming mitigation in multiple-input multiple-output (MIMO) communication systems has been studied in many existing works. These methods generally rely on estimating the spatial characteristics of the jammer and dedicating some of the degrees of freedom of the multiple-antenna receiver to mitigate its effect \cite{pirayesh2022jamming}, see for example \cite{marti2023universal}. This is not applicable in a pure OtA computation setting with a single-antenna receiver/fusion center.

In this paper, we propose and evaluate a detection scheme against active attacks perpetrated by an external user in pure OtA computation systems which, to the best of our knowledge, has not been investigated previously. For detecting the presence of the attacker, in every communication period, legitimate users send some dummy samples in addition to the real data, while the structure or statistics of the superimposed dummy samples can be used by the server for detector design. The dummy samples are set to zero and the composite data vector is hidden in a secret subspace, known only to the legitimate users and server, through multiplication with a shared unitary matrix. Then, after the server applies the inverse of the shared unitary, the attacker can be detected among the channel noise at the dummy sample positions of the received vector. We compare this scheme to the case where the dummy samples at different users are generated independently and demonstrate its superiority in both detection performance and channel resource usage.

%% file: OtA.tex
\section{Over-the-Air Computation over Fading Channels}
OtA computation takes advantage of the wave superposition property of wireless multiple-access channels to perform data aggregation in a communication-efficient manner. For this to work, multiple users transmit their data simultaneously after linear processing with amplitude modulation. The aggregated signals are post-processed at the server to achieve the intended function computation.

Assume that there are \numusers{} users transmitting data symbols $\vect{s} = {\left\{ s_k \in \complexes \right\}_{k = 1}^\numusers}$
to a server over a wireless fading channel with the channel coefficients ${\left\{ \channelcoef_k \in \complexes \right\}_{k = 1}^\numusers}$. The server obtains the estimated function value 
\begin{equation}
	f(\vect{s}) = \psi \left( \sum_{k=1}^{\numusers{}}\channelcoef_k\phi_n(s_k) + z \right),
\end{equation}
where ${\left\{\phi_{k}(\cdot)\right\}_{k = 1}^\numusers}$  are the pre-processing functions at the users, $\psi(\cdot)$ is the post-processing function at the server, and $z$ is additive white Gaussian noise with variance $\cnvar$. Through designing the pre- and post-processing functions, OtA computation can be used to compute a wide variety of aggregation functions such as the arithmetic mean outlined below \cite{OtAfunctions2009,OtAfunctions2013}.

With per-user maximum power constraint $P_0 \in \reals^+$, using 
\begin{equation}
	\psi(y) = \frac{y}{\eta}
\end{equation}
and
\begin{equation}
	\phi_n(s_k) = s_k\frac{\eta}{\numusers \channelcoef_k}, \forall k \in 1,2,...\numusers{}
\end{equation}
with amplitude-scaling factor 
\begin{equation}
	\psf = \sqrt[]{P_0} \min_k \left( \frac{\numusers\abs{\channelcoef_k}}{\abs{s_k}} \right),
\end{equation} 
we get 
\begin{equation}
	f(\vect{s}) = \frac{1}{\psf}\left( \sum_{k=1}^{\numusers{}}\channelcoef_k s_k\frac{\psf}{\numusers \channelcoef_k} + z\right) = \frac{1}{\numusers}\sum_{k=1}^{\numusers{}}s_k + \frac{z}{\psf},
\end{equation} 
which is the arithmetic mean of the symbols \vect{s} plus the effective channel noise with power $\frac{\cnvar{}}{\psf^2}$.

One major drawback of OtA computation is that it is very sensitive to adversarial attacks due to the lack of an error protection mechanism. Since all signals are aggregated in the air, there is no possibility for the server to identify individual outliers. This allows malicious users to directly affect the aggregation result. Furthermore, it is challenging for the server to determine whether the aggregation result is corrupted by perturbations from malicious attackers. Thus, we propose a scheme where users transmit some additional dummy data with known characteristics, to be used for detecting the presence of active attackers at the server side.

%% file: system_model.tex
\section{System Model}

\begin{figure}
	\centering
	\includegraphics[scale=0.1]{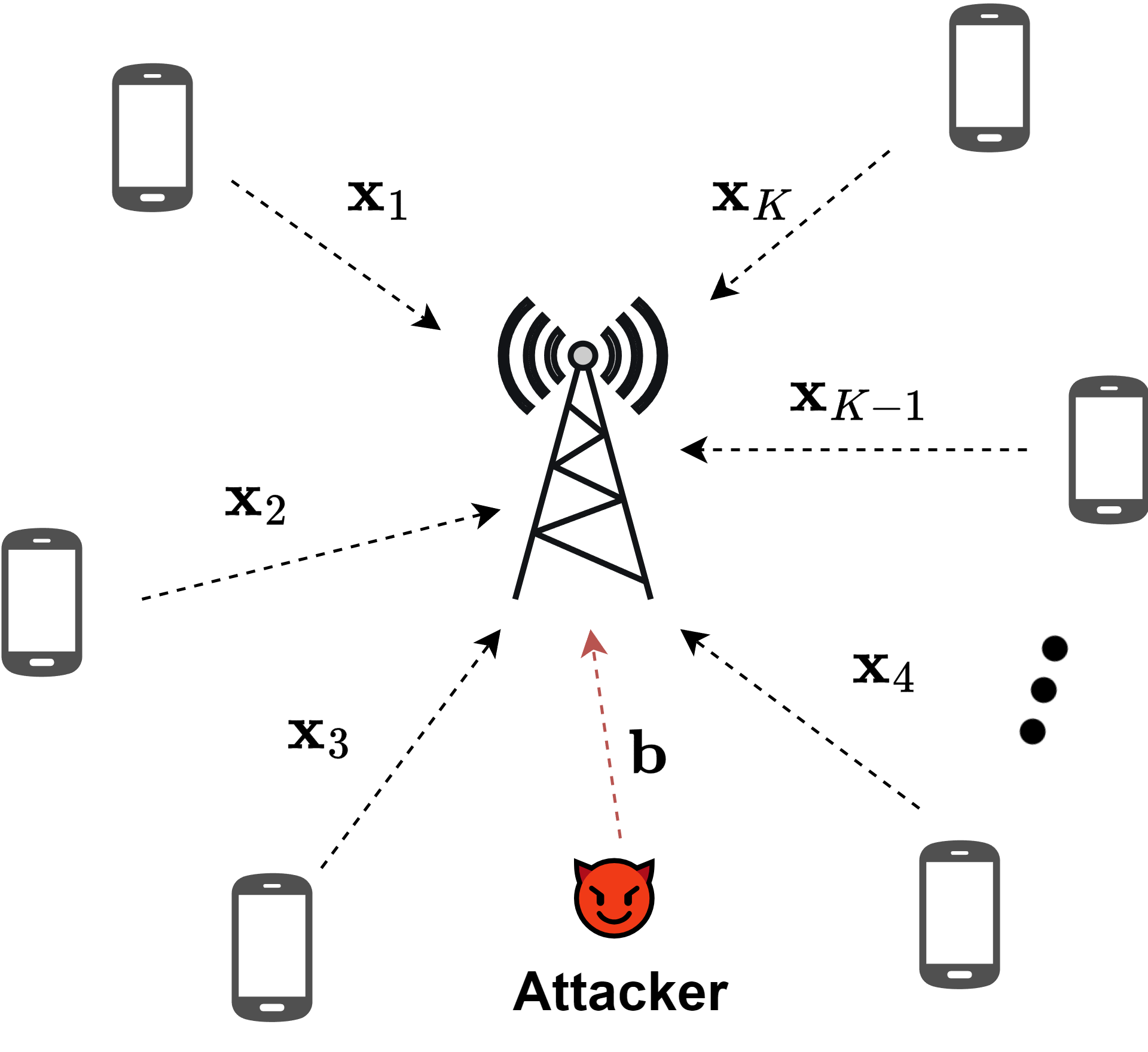}
	\caption{System consisting of K authentic users and 1 active attacker.}
	\label{fig:system_model}
\end{figure}

We consider an OtA computation system with \numusers{} legitimate users, one server, and one external attacker. The system is illustrated in Fig. \ref{fig:system_model}. Each legitimate user has a length-$L$ data vector to transmit. Ideally, the receiver needs to compute the arithmetic mean of the data vectors from all legitimate users as accurately as possible.

To detect the presence of malicious attackers in the system, we propose the following detection scheme using dummy samples. We divide time into equal-length blocks where each block contains a certain number of symbols (or channel uses). All channels are considered to remain static within a block but can vary across different blocks. The system activity in every block is divided into two phases: the communication phase and the detection phase. In each block, a legitimate user $k\in\{1,\ldots, \numusers\}$ transmits a composite data vector  $\datavector_k \in \complexes^{\numreal + \numdummy}$, consisting of \numreal{} real data symbols $\mathbf{x}_{k,r}  \in \complexes^{\numreal}$ in the communication phase and \numdummy{} dummy samples $\mathbf{x}_{k,d}  \in \complexes^{\numdummy}$ in the detection phase. The matrix forms of the composite data $\mathbf{X}$ and dummy samples  $\mathbf{X_D}$ are written as
\begin{align}
	\mathbf{X} &= 
	\begin{bmatrix}
		\vert & \vert &  & \vert\\
		\datavector_1   & 
		\datavector_2   &
		\dots & 
		\datavector_\numusers  \\
		\vert & \vert &  & \vert
	\end{bmatrix} \in \complexes^{(\numreal + \numdummy) \times \numusers },	\\ 
	\mathbf{X}_D &= 
	\begin{bmatrix}
		\vert & \vert &  & \vert\\
		\datavector_{1,d}   & 
		\datavector_{2,d}  &
		\dots & 
		\datavector_{\numusers,d}  \\
		\vert & \vert &  & \vert
	\end{bmatrix} \in \complexes^{\numdummy \times \numusers}.
\end{align}

We assume that the attacker has no knowledge of the existence of dummy samples and the detection scheme being used at the server. It transmits the perturbation signal $\byzdatavector \in \complexes^{\numreal + \numdummy}$, wherein $\byzdatavector_d \in \complexes^{\numdummy}$ is the perturbation signal vector during the detection phase.

Based on the received vector $\mathbf{y}_d=[y_1,\ldots, y_D]\in \complexes^\numdummy$ at the server during the detection phase, we can formulate the detection problem as a binary hypothesis test. Let $\channelcoef{}_k \in \complexes$ be the channel coefficient of user $k$, $h_b \in \complexes$ be the channel coefficient of the attacker  and  $\noise_d\sim \CN{0}{\cnvar \mathbf{I}_{D}}$ be the noise vector. The two hypotheses are:

\begin{equation}
	\begin{split}
		\mathcal{H}_0: \mathbf{y}_d&=  \frac{1}{\numusers}\sum_{k=1}^{\numusers} \mathbf{x}_{k,d} + \frac{\noise_d}{\psf}
		= \frac{1}{\numusers}\mathbf{X}_D\ones_{\numusers} + \frac{1}{\psf}\noise_d \\
		\mathcal{H}_1: \mathbf{y}_d	&=  \frac{1}{\numusers}\sum_{k=1}^{\numusers} \mathbf{x}_{k,d}  + \frac{\channelcoef_b \mathbf{b}_{d}}{\psf} + \frac{\noise_d}{\psf}\\
		&= \frac{1}{\numusers}\mathbf{X}_D\ones_{\numusers}
		+ \frac{\channelcoef_b}{\psf}\mathbf{b}_{d} 
		+ \frac{1}{\psf}\noise_d
	\end{split}
\end{equation}

where $\ones_{\numusers}$ is a length $\numusers$ column vector of all ones. 

\section{Dummy Sample Design and Detection Scheme}
The general framework makes no assumptions on the design of the dummy samples or the actual detection scheme. In this paper, we study two different scenarios: 1) the legitimate users and the server have access to some shared randomness (the \textit{correlated} case) and 2) the dummy samples are independently generated at the legitimate users (the \textit{uncorrelated} case). We describe both schemes and demonstrate the trade-off between detection performance and communication cost under a limited energy budget per block transmission.

\subsection{Uncorrelated Design}
In this scenario, the dummy samples at different users are generated independently of each other. Furthermore, the dummy sample generation mechanism should be chosen such that the attacker cannot perceive the difference between the communication and detection phases. Otherwise, the attacker could cease transmission to avoid being detected. Consequently, it is not possible for the legitimate users to simply stay idle during the detection phase. To obscure the existence of the detection phase, we consider that the dummy samples are randomly interleaved into the authentic data, i.e., the \numdummy{} indices are uniformly picked from  $\{1,2,...,\numreal + \numdummy\}$ without replacement. 

As an example, we consider that the dummy samples follow a zero-mean Gaussian distribution, $\mathbf{X}_D \sim \N{0}{\dnvar\mathbf{I}_{\numdummy \times \numusers}}$. Then, the hypothesis test becomes 
\begin{equation}
	\begin{split}
		&\mathcal{H}_0: \mathbf{y}_d=  \tilde{\mathbf{z}}\\
		&\mathcal{H}_1: \mathbf{y}_d= \frac{\channelcoef_b}{\psf}\mathbf{b}_{d} + \tilde{\mathbf{z}}
	\end{split}
\end{equation}

where $\tilde{\mathbf{z}} \sim \CN{0}{(\frac{\cnvar}{\eta^2} + \frac{\dnvar}{\numusers})\mathbf{I}_{\numdummy}}$ is the effective noise resulting from both the channel noise and the contributions from legitimate users.

Assume that each user has a fixed power budget proportional to the number of authentic data samples that are to be transmitted, i.e. $P_0 \propto \numreal$. Further, we assume that the power budget is uniformly allocated to the $\numreal + \numdummy$ transmitted samples\footnote{Non-uniform allocation is one possible extension that could be considered in future works.}. Then, the introduction of dummy samples reduces the per-symbol energy $P_s$ by a factor of $\frac{\numreal}{\numreal+\numdummy}$. This reduction applies to both dummy samples and true data. Consequently, while a large \numdummy{} might improve the detection performance, it also degrades the aggregation accuracy of the authentic data. Furthermore, the number of symbols that need to be transmitted over the channel increases by a factor of $\frac{\numreal+\numdummy}{\numreal}$, incurring an increased communication cost.

An additional concern exists in the case where the attacker has non-uniform energy distribution across the $\numreal + \numdummy$ samples. For example, if the attacker stays idle during the detection phase, it would be impossible to detect its presence. Depending on the underlying system and aggregation function, it might be more or less viable for the attacker to adopt such a strategy, so it is difficult to obtain any general conclusion on the detection performance. In the next section, we show that (aside from the other advantages) the shared randomness among legitimate users allows us to transform any attack into an (in expectation) energy-uniform one.

\subsection{Correlated Design}

We now discuss an improved scheme in the case where the users and the server share some secret in the generated dummy sample design. Relying on this secret, the users generate a shared unitary matrix $\mathbf{U} \in \complexes^{(\numreal + \numdummy)\times (\numreal + \numdummy)}$ in a pseudo-random manner, which is \textit{Haar} (uniformly) distributed over the set of all $(\numreal + \numdummy)\times (\numreal + \numdummy)$ unitary matrices \cite{genHaar2007mezzadri}. This matrix is then used to hide $\mathbf{X}$ in some  $(\numreal + \numdummy)$-dimensional subspace which is unknown to the attacker. The transmitted data in the matrix form is written as 
\begin{equation}
	\mathbf{\tilde{X}} = \mathbf{U}\mathbf{X} \in \complexes^{(\numreal + \numdummy)\times (\numusers)}.
\end{equation}
This can be realized by letting each user $k$ compute their composite data vector as $\tilde{\datavector}_k = \mathbf{U} \datavector_k$. The transmitted data matrix now consists of correlated data values, as opposed to the uncorrelated design. The server receives 
\begin{equation}
	\tilde{\mathbf{y}} =  \frac{1}{\numusers}\mathbf{U}\mathbf{X}\ones_{\numusers} + \frac{\channelcoef_b}{\psf} \mathbf{b} + \frac{1}{\psf}\noise,
\end{equation}
which is subsequently projected back to the original subspace and gives
\begin{equation}
	\mathbf{y} = \mathbf{U}^\dag\mathbf{\tilde{y}} = \frac{1}{\numusers}\mathbf{X}\ones_{\numusers} + \frac{\channelcoef_b}{\psf}\mathbf{U}^\dag \mathbf{b} + \frac{1}{\psf}\mathbf{U}^\dag\noise,
\end{equation}
where  $\dag$ notates the conjugate transpose and  $\noise\sim \CN{0}{\cnvar \mathbf{I}_{L+D}}$ is the channel noise.

One key advantage compared to the previous case is that we can let $\mathbf{X}_D=\mathbf{0}$ (no decrease in per-symbol energy) without alerting the attacker since the transmitted signal is hidden in a shared secret subspace. Although we do not allocate any energy to the dummy samples, there is still a cost of increased channel uses. With $\mathbf{X}_D=\mathbf{0}$, the hypothesis test reduces to 
\begin{equation}
	\begin{split}
		&\mathcal{H}_0: \mathbf{y}_d=  \tilde{\mathbf{z}}\\ 
		&\mathcal{H}_1: \mathbf{y}_d= \frac{\channelcoef_b}{\psf} \mathbf{U}_D^\dag\mathbf{b} + \tilde{\mathbf{z}},
	\end{split}
\end{equation}
where $\mathbf{U}_D$ consists of the \numdummy{} columns of $\mathbf{U}$ that correspond to the dummy samples and $\tilde{\mathbf{z}}=\frac{1}{\psf}\mathbf{U}_D^\dag\noise$. Since $\mathbf{U}$ is a unitary matrix, its columns form an orthonormal basis of $\complexes^{\numreal + \numdummy}$. As a result, the covariance of $\noise$ is preserved during the multiplication and thus $\tilde{\mathbf{z}} \sim \CN{0}{\frac{\cnvar}{\psf^2}\mathbf{I}_{\numdummy}}$. Furthermore, since $\mathbf{U}$ is Haar distributed, for any deterministic $\mathbf{b}$ we have
\begin{equation}
	\EX{\mathbf{u}_i^\dag \mathbf{b}}^2 = \frac{\norm{\mathbf{b}}^2}{\numreal + \numdummy} \, , \; i=1,2,...,\numreal+\numdummy,
\end{equation}
where $\mathbf{u}_i$ notates the $i$-th column of $\mathbf{U}$. That is, the energy of $\mathbf{U}^\dag\mathbf{b}$ is expected to be uniformly distributed across all its samples regardless of the distribution of $\mathbf{b}$.

\subsection{Detector}

With both the uncorrelated and correlated designs, the detection problem concerns detecting an unknown signal in noise (albeit with different signal and noise variance values). Let $\tilde{\sigma}^2$ denote the noise variance, with $\tilde{\sigma}^2=\frac{\cnvar}{\eta^2} + \frac{\dnvar}{\numusers}$ for the uncorrelated case and $\tilde{\sigma}^2=\frac{\cnvar}{\eta^2}$ for the correlated case. Due to the lack of knowledge about the attack signal distribution, we consider an energy detector with the detection rule 
\begin{equation}
	\norm{\mathbf{y}_d}^2 \decrule \gamma
\end{equation} 
for some threshold $\gamma$. 

With hypothesis $\mathcal{H}_0$, since each component of the received signal $y_i$ follows a zero-mean complex Gaussian distribution with variance $\tilde{\sigma}^2$ for all $i=1,\ldots, D$, we have 
\begin{equation}
	|y_i|^2 \sim \text{Exp}\left( \frac{1}{\tilde{\sigma}^2} \right) \, , \, i=1,\ldots, \numdummy.
\end{equation}
With $\norm{\mathbf{y}_d}^2=\sum_{i=1}^{D}|y_i|^2$, we have
\begin{equation}
	\label{eq:H0}
	\begin{aligned}
		\EX{\norm{\mathbf{y}_d}^2 \vert \; \mathcal{H}_0} &= 	\numdummy \tilde{\sigma}^2 \\
		\Var{\norm{\mathbf{y}_d}^2 \vert \; \mathcal{H}_0} &= 	\numdummy \tilde{\sigma}^4.
	\end{aligned}
\end{equation}

With hypothesis $\mathcal{H}_1$, the distribution of $\norm{\mathbf{y}_d}^2$ will depend on the perturbation signals.
In a special case, when the attack signals are generated according to a complex Gaussian distribution $\byzdatavector \sim \CN{0}{\mathit{\mathbf{I}}_{\numreal + \numdummy{}}}$, the components of $\mathbf{y}_d$ under $\mathcal{H}_1$, conditioned on $h_b$, are independent complex Gaussian variables. Let $y_i$ represent the $i$-th component of $\mathbf{y}_d$, then

\begin{equation}
	\label{eq:yiH1}
	|y_i|^2 \, \vert h_b \sim \text{Exp}\left( \frac{1}{\tilde{\sigma}^2 + \vert h_b \vert ^2/\eta^2} \right) \, , \, i=1,\ldots, \numdummy.
\end{equation}

From  \eqref{eq:yiH1} it follows that 
\begin{equation}
	\begin{aligned}
		\EX{|y_i|^2 \, \vert h_b,\mathcal{H}_1} &= 	 \tilde{\sigma}^2 + \vert h_b \vert ^2/\eta^2 \\
		\Var{|y_i|^2 \, \vert h_b,\mathcal{H}_1} &= \left( \tilde{\sigma}^2 + \vert h_b \vert ^2/\eta^2\right)^2.
	\end{aligned}
\end{equation}
Based on these, we can obtain the conditional mean and variance of $\norm{\mathbf{y}_d}^2$ under $\mathcal{H}_1$ as 
\begin{equation}
	\label{eq:H1_cond}
	\begin{aligned}
		\EX{\norm{\mathbf{y}_d}^2 \, \vert h_b,\mathcal{H}_1} &= 	 \numdummy \left( \tilde{\sigma}^2 + \vert h_b \vert ^2/\eta^2 \right)\\
		\Var{\norm{\mathbf{y}_d}^2 \, \vert h_b,\mathcal{H}_1} &= \numdummy \left( \tilde{\sigma}^2 + \vert h_b \vert ^2/\eta^2\right)^2.
	\end{aligned}
\end{equation}
Now, we can compute the mean of $\norm{\mathbf{y}_d}^2$ under $\mathcal{H}_1$ as
\begin{equation}
	\label{eq:H1_mean}
	\begin{aligned}
		\EX{\norm{\mathbf{y}_d}^2 \, \vert \mathcal{H}_1} &= \mathbb{E}_{h_b}\left[\EX{\norm{\mathbf{y}_d}^2 \, \vert h_b,\mathcal{H}_1} \right] \\
		&= \numdummy \left( \tilde{\sigma}^2 + \beta / \eta^2 \right),
	\end{aligned}
\end{equation}
where $\beta=\EX{\vert h_b \vert ^2}$ is the average path gain (in power) of the attacker, which is assumed to be constant during the detection phase. The variance can be computed using the conditional law of total variance:
\begin{align}
	\label{eq:H1_var}
	\Var{\norm{\mathbf{y}_d}^2 \, \vert \mathcal{H}_1} &= \mathbb{E}_{h_b}\left[ \Var{\norm{\mathbf{y}_d}^2\vert h_b, \mathcal{H}_1}\right]  \\ &+ \text{Var}_{h_b} \left[ \EX{\norm{\mathbf{y}_d}^2 \vert h_b, \mathcal{H}_1} \right] \nonumber \\
		&= \numdummy \left( \tilde{\sigma}^4 + \frac{2\tilde{\sigma}^2 \beta}{\eta^2} + \frac{2\beta^2}{\eta^4} \right) + \numdummy^2 \frac{\beta^2}{\eta^4}, \nonumber
\end{align}
where the last equality uses the results from \eqref{eq:H1_cond} and that $\vert h_b \vert ^2$ is exponentially distributed. We now have the mean and variance of $\norm{\mathbf{y}_d}^2$ under both hypotheses, given in \eqref{eq:H0}, \eqref{eq:H1_mean} and \eqref{eq:H1_var}. With the correlated design, we have $\tilde{\sigma}^2=\frac{\cnvar}{\eta^2}$, which gives
\begin{equation}
	\label{eq:all_correlated}
	\begin{split}
		\EX{\norm{\mathbf{y}_d}^2 \vert \; \mathcal{H}_0} &= 	\numdummy \frac{\sigma^2}{\eta^2} \\
		\Var{\norm{\mathbf{y}_d}^2 \vert \; \mathcal{H}_0} &= 	\numdummy \frac{\sigma^4}{\eta^4} \\
		\EX{\norm{\mathbf{y}_d}^2 \, \vert \mathcal{H}_1} &= \numdummy \frac{\sigma^2 + \beta}{\eta^2} \\
		\Var{\norm{\mathbf{y}_d}^2 \, \vert \mathcal{H}_1} &=  \frac{1}{\eta^4} \left[ \numdummy \left( \sigma^4 + 2\sigma^2\beta + 2\beta^2 \right) + \numdummy^2\beta^2 \right].
	\end{split}
\end{equation}
The energy detector is expected to perform well if the difference in expected values between the two hypotheses is large compared to the standard deviations. The probability of detection is high when 
 \begin{align}
	 	\frac{\EX{\norm{\mathbf{y}_d}^2 \, \vert \mathcal{H}_1} - \EX{\norm{\mathbf{y}_d}^2 \, \vert \mathcal{H}_0}}{\Var{\norm{\mathbf{y}_d}^2 \, \vert \mathcal{H}_1}} \\ = \sqrt{\frac{1}{1+\frac{1}{D}\left( \frac{\sigma^4}{\beta^2} + \frac{2\sigma^2}{\beta} + 2\right) }}, \nonumber
 \end{align}
 is large. This occurs for small channel noise, when using many dummy samples and/or when the attacker has good channel conditions, which is in line with what intuition tells us. 

%% file: results.tex
\section{Numerical Results}

\begin{figure}[t!]
	\centering  
	\subfigure[Uncorrelated design.] {\label{fig:ROC_uncorrelated}\includegraphics[width=0.9\columnwidth]{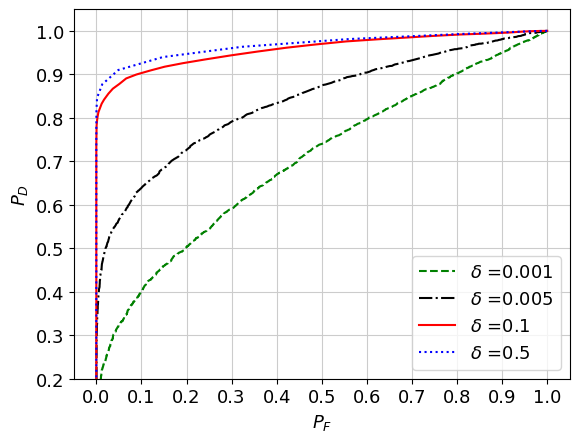}}
	\subfigure[Correlated design.]{ \label{fig:ROC_correlated}\includegraphics[width=0.9\columnwidth]{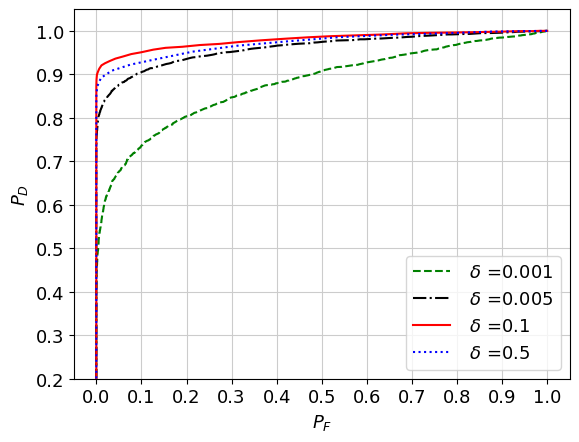}}\\
	\caption{Receiver operator characteristics for systems consisting of various number of dummy samples.}
	\label{fig:ROC}
\end{figure}

\begin{figure}[ht!]
	\centering  
	\subfigure[Shared maximum energy constraint.]{ \label{fig:main_normal}\includegraphics[width=0.9\columnwidth]{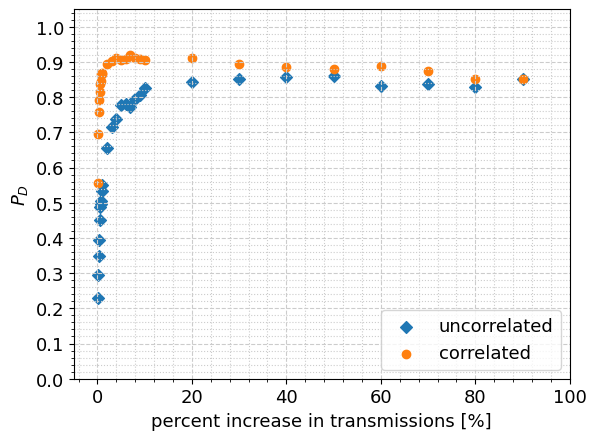}} \\
	\subfigure[Legitimate users can use more energy than the attacker.]  {\label{fig:main_power}\includegraphics[width=0.9\columnwidth]{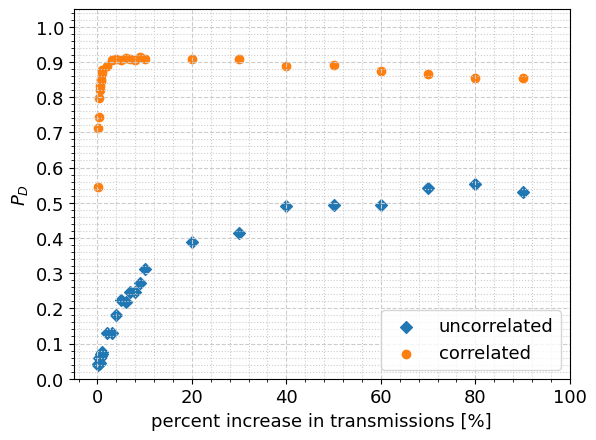}}
	\caption{Trade-off between the probability of detection and the increase in data transmission for fixed $P_F=0.01$, depending on the access to shared randomness. }
	\label{fig:main}
\end{figure}

\begin{figure*}
	\centering  
	\subfigure[$\delta = 0.01$]{ \label{fig:overlap1}\includegraphics[width=0.66\columnwidth]{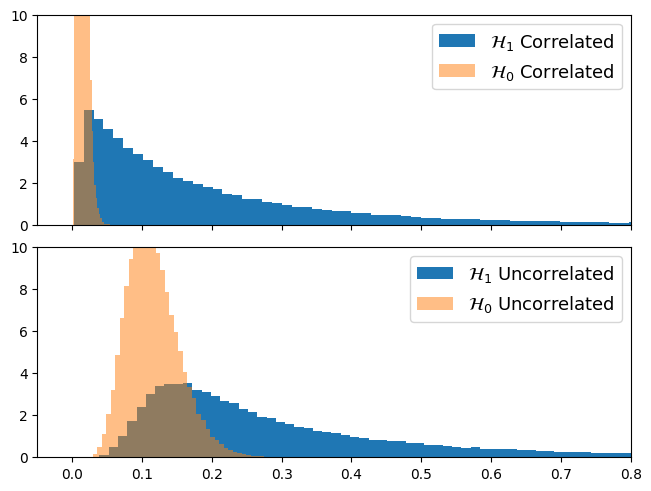}}
	\subfigure[$\delta = 0.1$] {\label{fig:overlap2}\includegraphics[width=0.66\columnwidth]{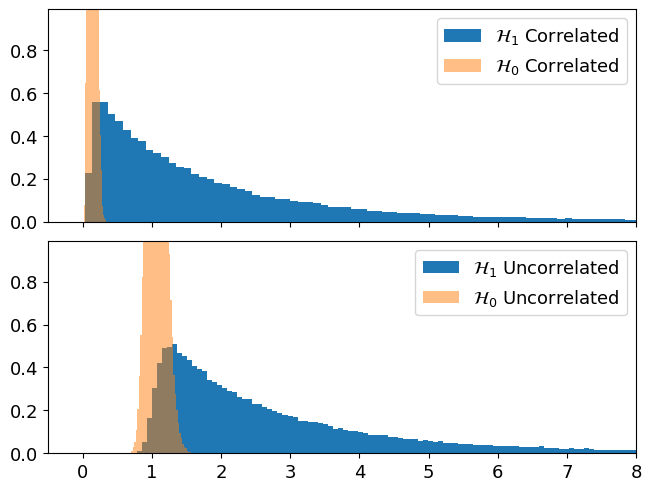}}
	\subfigure[$\delta = 0.5$] {\label{fig:overlap3}\includegraphics[width=0.66\columnwidth]{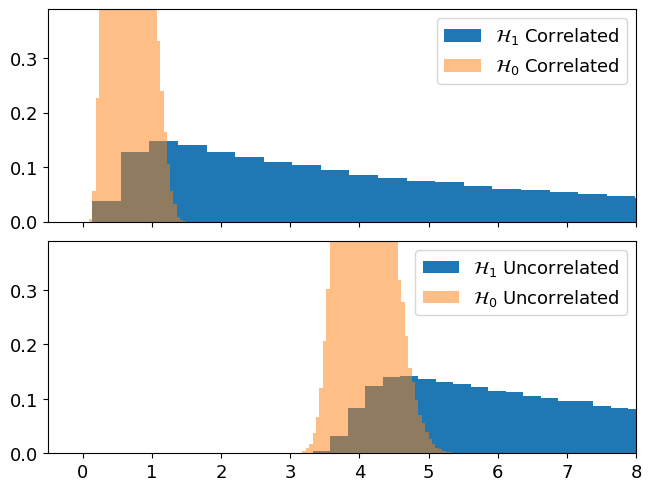}}
	\caption{Comparison of the distinguishability of hypothesis for the two different systems, for various $\delta$ values.}
	\label{fig:overlap}
\end{figure*}
In this section, we evaluate and compare the performance of the two dummy sample designs. Our system consists of $\numusers = 100$ legitimate users, $1$ server and $1$ attacker. The users are randomly positioned in a circular area with radius $100$m, centered around the server. Each legitimate user transmits $\numreal = 1000$ authentic data samples and $\numdummy$ dummy samples with the ratio given as $\delta=D/L$. The legitimate users and the attacker have a total power budget $P_0 = 1$mW. The channel coefficients comprise two factors, small- and large-scale fading. The small-scale fading follows $\CN{0}{1}$, and the large-scale fading is considered as $d^{-4}$ where $d$ is the link distance. We include an additional step to limit the effective noise by only allowing users with small-scale fading above $0.2$ to participate in the aggregation. The channel noise variance is $\cnvar=-110$dBm. The simulations are performed with uniform-energy attack, which means that the attacker's energy is uniformly spread within a block.
Specifically, we consider a random complex Gaussian attacker vector $\byzdatavector \sim \CN{0}{\mathit{\mathbf{I}}_{\numreal{}+\numdummy{}}}$ that is transmitted at maximum power.

Fig. \ref{fig:ROC} shows the receiver operator characteristic (ROC) of the two different designs. We observe clear advantages of the correlated design in those scenarios with fewer dummy samples, indicating that it allows for a design with significantly less impact on the amount of data transmissions. We see that $\delta = 0.005$ for the correlated design achieves a similar performance to $\delta = 0.1$ in the uncorrelated design, with $20$ times less dummy samples. This difference is further illustrated in Fig. \ref{fig:overlap}, which shows how the distinguishability between the hypotheses evolves as $\delta$ increases. The plots show histograms of the received energies $\norm{\mathbf{y}_d}^2$ from $100000$ realizations of the system. In Fig. \ref{fig:overlap1}, we notice that it is significantly easier to distinguish between the two hypotheses for the correlated design (the upper plot) than the uncorrelated one (the lower plot). As $\delta$ increases from $0.01$ to $0.1$ in Fig. \ref{fig:overlap2} and then $0.5$ in Fig. \ref{fig:overlap3}, we observe that this difference diminishes, which is in line with the previous result.

Finally, Fig. \ref{fig:main} shows the probability of detection (for a fixed probability of false alarm) as a function of the additional channel uses incurred by transmitting the dummy samples. In Fig. \ref{fig:main_normal}, the attacker and legitimate users share the same maximum energy constraint (as outlined above). We note that our correlated design with access to shared randomness among legitimate users achieves very good detection performance with a very small amount of added resources (channel uses). While the uncorrelated design also reaches a considerable probability of detection eventually, this occurs with a significantly larger communication cost. The reason that the uncorrelated design can reach such a good detection performance (close to the correlated design) is because of the power control applied when performing OtA computation. All but one of the legitimate users decrease their transmission power based on the user with the worst channel condition so that all signals are received with equal power. The consequence of this is that the received attacker perturbation is non-negligible when compared to the part of the noise variance that stems from the $100$ legitimate users. Fig. \ref{fig:main_power} illustrates a scenario where legitimate users are allowed to use $10$ times more energy than the attacker. As expected, we now see that the performance of the uncorrelated design becomes much worse, while the correlated design is unaffected. This could occur if, for example, the attacker limits its power output in order to avoid being easily detected. Such behavior makes it harder for the server to detect when the dummy samples are uncorrelated.

%% file: conclusions.tex
\section{Conclusions}
In this paper, we propose a detection scheme for active attacks in OtA aggregation systems. The main idea is to, alongside the authentic data, transmit some additional dummy samples that can be used by the receiver for the detection task. Depending on the possibility of allowing shared randomness between the users and the server, we propose two different designs and analyze their detection performances. For the case with shared randomness, we show that we can achieve good detection performance for only a small increase in data transmissions, without degrading the aggregation accuracy of the authentic data.